\documentclass[aps,pra,reprint,superscriptaddress,longbibliography]{revtex4-1}
\usepackage{graphicx}
\usepackage{psfrag, color}
\usepackage{verbatim}
\usepackage{amsmath}

\begin{document}

\title{Impact of Heterostructure Design on Transport Properties in the Second Landau Level of \emph{in-situ} Back-Gated Two-Dimensional Electron Gases}


\author{J. D. Watson} \altaffiliation[Present Address:]{Kavli Institute of Nanoscience, Delft University of Technology, 2600 GA Delft, The Netherlands.}
\affiliation{Department of Physics and Astronomy, Purdue University, West Lafayette, IN 47907, USA}
\affiliation{Birck Nanotechnology Center, Purdue University, West Lafayette, IN 47907, USA}
\author{G. A. Cs\'{a}thy}
\affiliation{Department of Physics and Astronomy, Purdue University, West Lafayette, IN 47907, USA}
\author{M. J. Manfra}
\email{mmanfra@purdue.edu}
\affiliation{Department of Physics and Astronomy, Purdue University, West Lafayette, IN 47907, USA}
\affiliation{Birck Nanotechnology Center, Purdue University, West Lafayette, IN 47907, USA}
\affiliation{School of Electrical and Computer Engineering, Purdue University, West Lafayette, IN 47907, USA}
\affiliation{School of Materials Engineering, Purdue University, West Lafayette, IN 47907, USA}

\begin{abstract}
We report on transport in the second Landau level in \emph{in-situ} back-gated two-dimensional electron gases in GaAs/Al$_x$Ga$_{1-x}$As quantum wells. Minimization of gate leakage is the primary heterostructure design consideration. Leakage currents resulting in dissipation as small as $\sim$ 10 pW can cause noticeable heating of the electrons at 10 mK, limiting the formation of novel correlated states.  We show that when the heterostructure design is properly optimized, gate voltages as large as 4V can be applied with negligible gate leakage, allowing the density to be tuned over a large range from depletion to over 4 $\times$ 10$^{11}$ cm$^{-2}$.  As a result, the strength of the $\nu = 5/2$ state can be continuously tuned from onset at n $\sim 1.2 \times 10^{11}$ cm$^{-2}$ to a maximum $\Delta_{5/2} = 625$ mK at n = $3.35 \times 10^{11}$ cm$^{-2}$.  An unusual evolution of the reentrant integer quantum Hall states as a function of density is also reported.  These devices can be expected to be useful in experiments aimed at proving the existence of non-Abelian phases useful for topological quantum computation.
\end{abstract}

\maketitle
\section{Introduction}

Since the discovery of the fractional quantum Hall effect (FQHE) at $\nu = 5/2$ \cite{willett1987observation}, this state has drawn intense scrutiny.  Much of the motivation for the study of the $\nu = 5/2$ state comes from the fact that numerical work \cite{morf1998transition,rezayi2000incompressible,peterson2008finite,storni2010fractional} has shown strong overlaps with the Pfaffian wavefunction \cite{moore1991nonabelions} and its particle-hole conjugate state, the so-called anti-Pfaffian \cite{lee2007particle,levin2007particle}, both of which are non-Abelian and could find uses in topologically protected quantum computation  \cite{kitaev2003fault,dassarma2005topologically,nayak2008non-abelian}.  
There have also, however, been theoretically proposed wavefunctions for the $\nu = 5/2$ state that exhibit Abelian statistics (see \cite{bishara2009interferometric,yang2013influence} for a summary of candidate states).  To date, the experimental tests to determine the nature of the ground state at $\nu = 5/2$ have failed to agree on the identity of the wavefunction.  Experiments probing the temperature dependence of tunneling between the edge states at $\nu = 5/2$ have been proposed \cite{fendley1995exact} and conducted \cite{radu2008quasiparticle,lin2012measurements} as a way to measure the quasiparticle effective charge $e^*$ and Luttinger liquid interaction parameter $g$ in order to discriminate between proposed wavefunctions.  These experiments, however, were inconclusive as tunneling experiments performed on the same Hall bar mesa but with different electrostatic confinement potentials gave results consistent with the non-Abelian anti-Pfaffian and $U(1) \times SU_2(2)$ states \cite{radu2008quasiparticle} and the Abelian 331 state \cite{lin2012measurements}.  Later experients by a different group \cite{baer2014experimental} were most consistent with the 331 state, but recent measurements of the spin polarization at $\nu = 5/2$ using NMR techniques \cite{tiemann2012unraveling,stern2012nmr}  indicate a fully spin-polarized electron gas which is inconsistent with the unpolarized 331 state \cite{halperin1983theory}.  Interferometry experiments showing an alternating Aharanov-Bohm period \cite{willett2009measurement,willett2010alternation}  are also consistent with a non-Abelian state at $\nu = 5/2$ and thus appear to rule out the 331 state.  As it is unclear how possible edge reconstruction \cite{chamon1994sharp} due to shallow confining potentials might influence the interpretation of the tunneling  experiments, it is possible that the different confinement parameters in the previously studied devices could be responsible for this apparent discrepancy.  

Given the complications associated with these experiments, it would therefore be desirable to examine transport in nanostructures in the quantum Hall regime in samples in which the electron density and confining potential could be tuned simultaneously in a single structure.  A variable density would also allow for direct comparisons between experimental results in the 2$^{\text{nd}}$ Landau level (LL) and the more well-understood lowest LL in a single device without the need for extremely large magnetic fields.  In addition, the original proposal \cite{chamon1997two} for an edge-state interferometer designed to directly measure the quantum statistics of the quasiparticles at $\nu= 5/2$ called for a device with a global back-gate to allow magnetic field sweeps at a constant filling fraction.  As such, there is strong motivation for back-gated devices exhibiting strong FQHE in the 2$^{\text{nd}}$ LL concomitant with a large range of density tuneability.

\section{Device Growth and Fabrication}
\begin{figure}[b]
\includegraphics[width=3in]{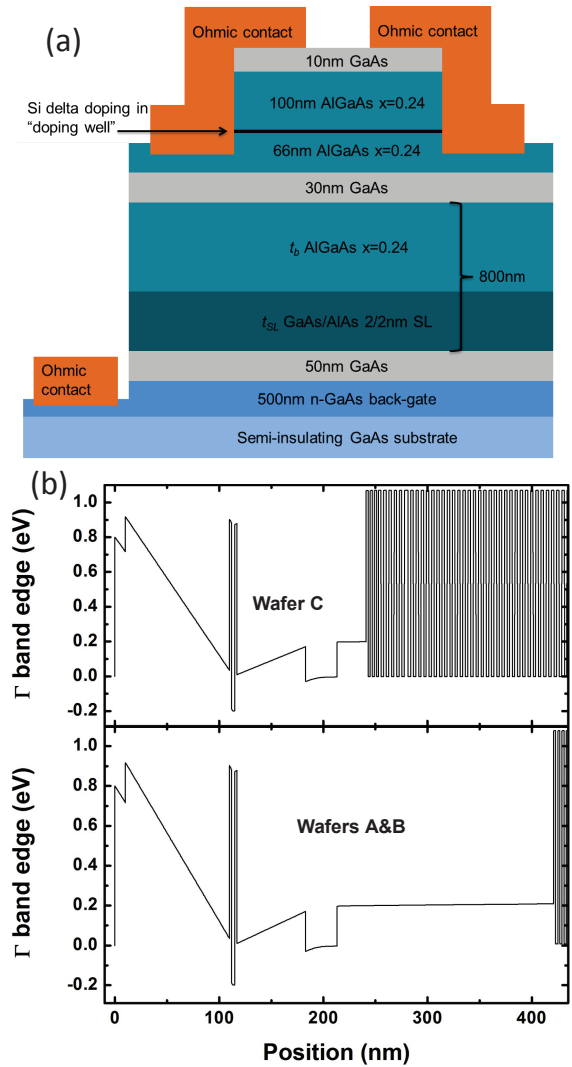}
\caption{(Color online). Heterostructure and device design. (a) Cross section of device showing heterostructure layer sequence and lithographic design. Dimensions are not to scale. Two separate etch steps define ``via'' holes to the backgate and the Van der Pauw mesa.  Ohmic contacts to the 2DEG and the gate are deposited in a single evaporation. TiAu pads (not shown for clarity) are deposited after the Ohmic contacts to facilitate wire bonding. (b) Self-consistent Schr\"{o}dinger/Poisson calculation of band structure showing the different position of the superlattice in each wafer.}
\label{Cross_Section}
\end{figure}

\begin{figure}[b]
\includegraphics[width=3in]{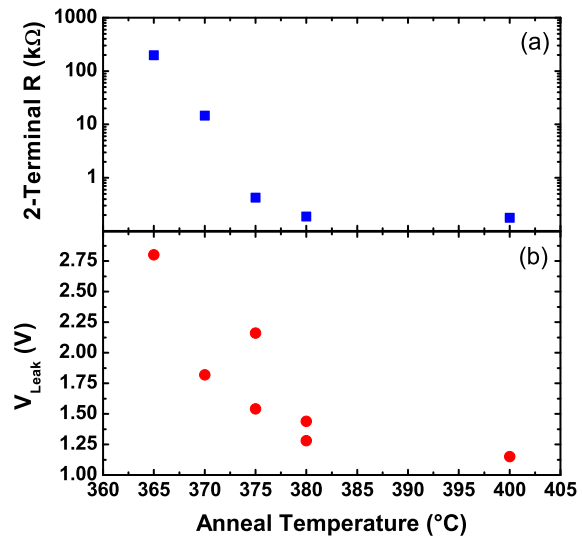}
\caption{(Color online). Effect of Ohmic annealing temperature on device performance from wafer A. (a) Median 2-terminal resistance to ground of individual contacts measured in the dark at T = 4 K and V$_{\text{g}}$ = 0 as a function of annealing temperature. (b) V$_{\text{leak}}$, defined as the voltage at which the gate leakage current reached 1 nA, as a function of annealing temperature. Two data points each are shown at 375 $^{\circ}$C and 380 $^{\circ}$C as this was near the optimal annealing temperature.}
\label{ContactR}
\end{figure}

In order to undertake such experiments, however, it is necessary to have a thorough understanding of how heterostructure design and device fabrication parameters affect device yields and the quality of transport in the 2$^{\text{nd}}$ LL.  Towards this aim, we have grown and processed a series of high quality, \emph{in-situ} backgated two-dimensional electron gases (2DEGs).  The processing of similar devices of lower mobility has been reported \cite{meirav1988high,Ritchie1991growth,hamilton1992backgated,hirayama1998twodimensional,kawaharazuka2000formation,muraki2000gaas,kawaharazuka2001free,valeille2008highly} and a similar high mobility device has been used to examine the energy gaps of FQHE states in the 2$^{\text{nd}}$ LL \cite{nuebler2010density}. However, to our knowledge there has not been a published systematic study of heterostructure design and processing conditions and their impact on the visibility of states in the 2$^{\text{nd}}$ LL.

 We studied three wafers utilizing two heterostructure designs summarized in Fig.~\ref{Cross_Section} to study the impact of the heterostructure on the gate leakage and the low temperature transport.  Both designs feature a 2DEG located approximately 200 nm from the surface in a 30 nm GaAs quantum well flanked by Al$_{0.24}$Ga$_{0.76}$As barriers modulation doped from the top side only at a setback of $\sim$ 70 nm.  The dopants were incorporated in a so-called doping well scheme (also known as a short-period superlattice) \cite{eisenstein2002insulating, pfeiffer2003therole, umansky2009mbe, manfra2014molecular} which has been found empirically to maximize the FQHE energy gaps in the 2$^{\text{nd}}$ LL. The \emph{in-situ} gate consisted of an n+ GaAs layer situated 850 nm below the bottom interface of the quantum well.  The key difference between the two designs was that in design \#1 an Al$_{0.24}$Ga$_{0.76}$As barrier of thickness $t_b = 200$ nm separated the quantum well from a GaAs/AlAs (2/2 nm) superlattice while in design \#2 $t_b$ was decreased to 20 nm while keeping the overall gate setback fixed at 850 nm.  Wafers A and B utilized design \#1 while wafer C utilized design \#2.

Device fabrication began by etching via holes to the gate layer using an etchant consisting of 50:5:1 water:phosphoric:peroxide followed by a second, $\sim$ 160 nm deep etch to define 1 mm Van der Pauw square mesas.  Ohmic contacts consisted of a 8/80/160/36 nm stack of Ni/Ge/Au/Ni and were annealed for 1 min in forming gas at a variety of temperatures.  Following the annealing, large TiAu pads off of the mesa were deposited in order to facilitate wirebonding.

Figure \ref{ContactR} shows the effect of annealing temperature on the quality of the contacts and the gate leakage measured in the dark at T = 4 K on devices fabricated from wafer B.  The lead resistance of the measurement setup was $\sim$ 1 $\Omega$, so the 2-terminal resistance values quoted here are reasonable proxies for the true contact resistance.  At an annealing temperature of 360 $^{\circ}$C, the contacts were electrically open at low temperature, and the contact morphology was extremely smooth, indicating that the metal did not melt or diffuse significantly during the anneal.   Figure \ref{ContactR}b displays the effect of the annealing on the gate leakage.  To quantify the leakage from our devices, we defined V$_{\text{leak}}$ as the gate voltage V$_{\text{g}}$ at which the gate leakage current reached 1 nA; thus high values of V$_{\text{leak}}$ are expected for a high quality gate insulating layer.  Both the 2-terminal resistance and V$_{\text{leak}}$ decrease monotonically as the annealing temperature is increased and the NiAuGe diffuses further into the semiconductor.  

To further study the impact of mask design and processing parameters on the gate leakage and contact resistance, we fabricated a set of test structures (not shown) which gave evidence that the gate leakage was primarily through the annealed contacts and not through the bulk of the mesa.  In addition, the test structures gave evidence that the leakage current density through annealed metal in etched regions was larger than that through annealed metal in un-etched regions.  The increased electric field due to the decreased gate-contact separation in the etched regions was insufficient to account for this increase in leakage density. This observation appears to imply that the etching procedure enhances the subsequent diffusion of the contacts.  With this in mind, we designed our lithographic mask sets to minimize the total Ohmic area, particularly in the region off of the etched mesa. In our final design the total Ohmic area was $<$ 1.5 $\times$ 10$^{4}$ $\mu$m$^{2}$ per device, and the total Ohmic overhang off each mesa was $\sim$ 6000 $\mu$m$^{2}$.  By minimizing the total time the etched sidewall of the mesa was exposed to air between the etch step and the metallization (typically $\sim$ 3-4 hours) and optimizing the geometry of the Ohmic contacts to include 45$^{\circ}$ scallops, we were able to produce devices with acceptably low contact resistances in the range of a few hundred Ohms while minimizing the gate leakage.
\begin{figure}[t]
\includegraphics[width=3in]{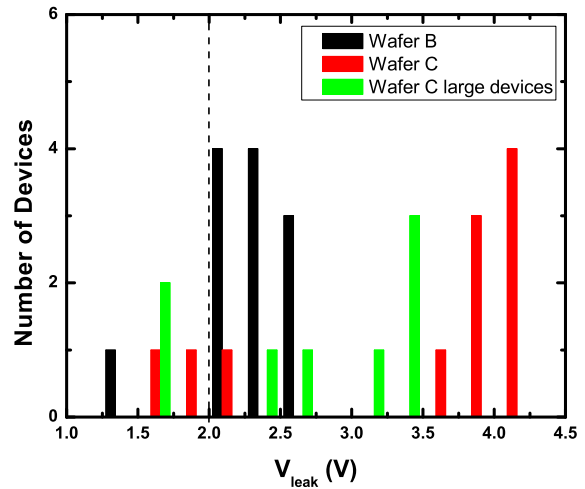}
\caption{(Color online). Histogram of leakage turn-on voltage V$_{\text{leak}}$ for devices fabricated with the optimized processing recipe and mask set.  All the devices were annealed at 375 $^{\circ}$C.  The dashed line represents the voltage required to reach a 2DEG density of $\sim$ 3.2 $\times$ 10$^{11}$ cm$^{-2}$.}
\label{histogram}
\end{figure}

Next, we examined the impact of heterostructure design on device performance.  Using our optimized fabrication recipe and mask set, we fabricated devices on both wafers B and C, using an annealing temperature of 375 $^{\circ}$C.  Figure \ref{histogram} is a histogram of the leakage turn-on  V$_{\text{leak}}$ for devices from each wafer.  The leakage in the majority of devices from wafer B (black bars) turned on around 2.2 V while the leakage in devices from wafer C (red bars) typically turned on around 3.8 V.  Evidently, the proximity of the superlattice to the quantum well has a pronounced effect on the gate leakage.  

The dashed line in Fig.~\ref{histogram} represents the gate voltage required in our geometry to reach a 2DEG density of $\sim$ 3.2 $\times$ 10$^{11}$ cm$^{-2}$, roughly twice the zero-bias density and the approximate electron density of 2DEGs exhibiting state-of-the-art energy gaps in the 2$^{\text{nd}}$ LL (see for instance references \cite{deng2014fractional,samani2014low,manfra2014molecular}).  As the devices from wafer C clearly could be biased well beyond the point necessary to study the FQHE of the 2$^{\text{nd}}$ LL, we fabricated Hall bar devices with larger contacts on wafer C to check how much less stringent the device design and fabrication requirements were for this wafer to exhibit acceptable gate leakage. These devices were based on a design \cite{miller2007fractional} known to both exhibit high quality transport in the 2$^{\text{nd}}$ LL and allow the incorporation of nanostructures.   The total Ohmic area per device was 3.0 $\times$ 10$^{5}$ $\mu$m$^2$ with 4.6 $\times$ 10$^4$ $\mu$m$^2$ overhanging the edge of the mesa.  Even though the Ohmic area in the etch field increased by a factor of $\sim$ 8 and the total Ohmic area increased by a factor of $\sim$ 20 from our optimized mask design, the leakage turn-on in most devices was still beyond 2.5 V, further highlighting the importance of proper heterostructure design.

We speculate that the large reduction in gate leakage in wafer C is due to two effects.  First, the alternating layers of the superlattice act as a diffusion barrier \cite{petroff1984impurity,schaff1984superlattice} to the metal from the Ohmic contacts; thus, by moving the superlattice closer to the quantum well, less metal is able to diffuse towards the gate, thereby reducing the shorting of the Ohmics to the gate.  In addition, Fowler-Nordheim tunneling \cite{lenzlinger1969fowler-nordheim,weinberg1982on,smoliner1987fowler,hickmott1984resonant,jensen2003electron} from the bulk of the 2DEG to the gate can be expected to be reduced by moving the tall AlAs barriers of the superlattice closer to the 2DEG.  

While moving the superlattice closer to the quantum well has the benefit of dramatically increasing the maximum achievable density, it also has the undesirable consequence of placing a significant amount of AlAs close to the quantum well.  It is known that Al is an effective getter of vacuum impurities during MBE growth \cite{pfeiffer2003therole}, and thus moving the superlattice closer to the 2DEG may degrade the quality of the FQHE states.  Indeed, the average maximum electron mobility in devices from wafer B was $\sim 15 \times 10^{6}$ cm$^2$/Vs while that from wafer C was $\sim 11 \times 10^{6}$ cm$^2$/Vs. Wafers B and C were grown on the same day, so it appears likely that the decrease in mobility can be attributed to the change in heterostructure design.  That being said, it has become clear in recent years that the zero field mobility is not a good predictor of energy gaps in the 2$^{\text{nd}}$ LL \cite{nuebler2010density,deng2014fractional,dassarma2014mobility,manfra2014molecular}.  Consequently, it was necessary to examine the magnetotransport at low temperature to make any definitive statement on the potential negative impact of moving the superlattice closer to the 2DEG.

\section{Impact of Heat Sinking and Heterostructure Design on RIQHE States}
\begin{figure}[t]
\includegraphics[width=3in]{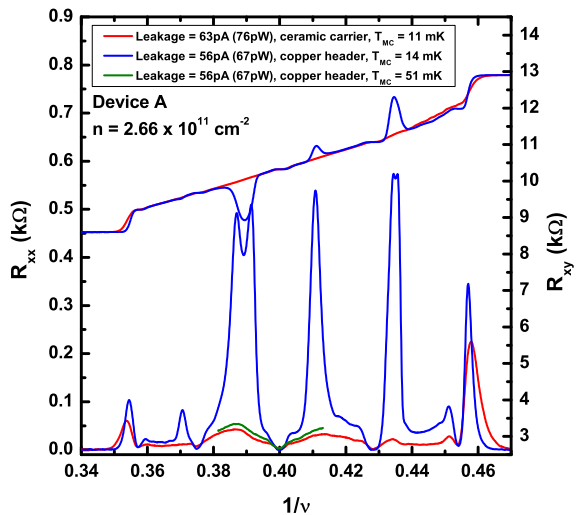}
\caption{(Color online). Magnetotransport in the lower spin branch of the 2$^{\text{nd}}$ LL in device A after illumination with a red LED.  During the first cool down of the sample (red curves) the sample was mounted on a commercial ceramic chip carrier.  At a gate leakage current (power) of $\sim$ 63 pA (76 pW) the electrons appear very warm as seen by the lack of RIQHE features, despite a low mixing chamber temperature T$_{\text{MC}}$.  During the second cool down of the sample (blue curves), the device was mounted on a homemade header with a copper strip screwed onto the end of the cold finger on the mixing chamber.  The electrons were obviously much colder even for a slightly higher T$_{\text{MC}}$.  The green curve shows the transport around $\nu = 5/2$ during the second cooldown for T$_{\text{MC}}$ = 51 mK.  Comparing the green and black data, we conclude that the electron temperature was $\sim$ 50 mK for T$_{\text{MC}}$ = 11 mK during the first cooldown.}
\label{cooling}
\end{figure}
Figure \ref{cooling} illustrates the importance of minimizing the gate leakage and properly heat sinking the sample in order to study the 2$^{\text{nd}}$ LL at low temperatures (T $< 25$ mK).  The data shown were taken from an early device from wafer A which was fabricated prior to the final optimization of our processing recipe.  During the first cool down of the device, the Joule heating of the electrons due to the gate leakage current evidently caused the electron temperature to depart from the mixing chamber temperature T$_{\text{MC}}$ for a gate leakage current (power) $\sim$ 4 pA ($\sim$ 3.5 pW) as evinced by weakening of the reentrant integer quantum Hall effect (RIQHE) features (data not shown).  By contrast, the excitation current of 2.1 nA contributed a neglible power dissipation of $\sim$ 45 fW at $\nu = 5/2$.  In order to facilitate wire bonding, we mounted the device on a commercial bondable ceramic chip carrier during the first cool-down.  This meant, however, that the sample was only cooled through the 18 $\mu$m thick Au bond wires.  To improve the heat sinking, we re-wired the same device on a homemade header.  In this design the sample was mounted to a strip of Cu with Ag paint, and the Cu strip was screwed directly onto the Cu cold finger of the mixing chamber resulting in a continuous metal connection between the mixing chamber and sample.  With this improved heat sinking, heating of the electrons was not evident until a gate leakage current (power) $\sim$ 56 pA (67pW).  Figure \ref{cooling} illustrates the vast improvement in electron temperature achieved by improving the heat sinking of the sample.  For a fixed density and approximately constant gate leakage current, the data taken with the Cu strip header show strong RIQHE features while the data taken with the ceramic chip carrier shows no RIQHE features.  To quantify T$_{\text{electron}}$ during the first cool-down, we show data (green curve in Fig.~\ref{cooling}) taken at T$_{\text{MC}}$ = 51 mK during the second cooldown.  The insulating peaks in R$_{\text{xx}}$ in the vicinity of $\nu = 5/2$ at T$_{\text{MC}}$ = 51 mK during the second cool-down are comparable to those seen at T$_{\text{MC}}$ = 11 mK during the first cool-down.  This allows us to estimate T$_{\text{electron}} \sim$ 50 mK for the black curves in Fig.~\ref{cooling}.


\begin{figure}[t]
\includegraphics[width=3in]{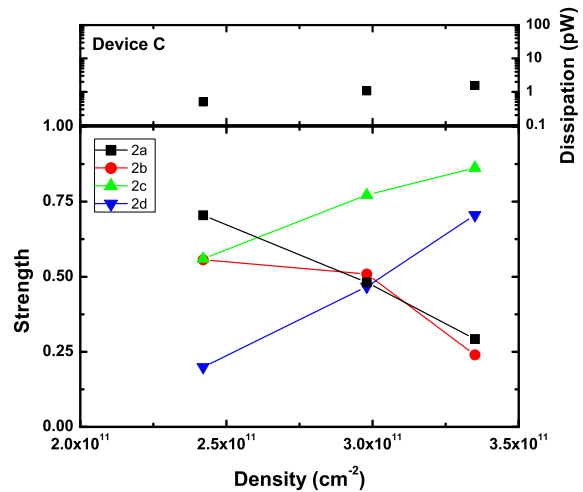}
\caption{(Color online). Strength (as defined in the text) of the RIQHE in device C during the second cool-down as a function of density; the power dissipation from the gate leakage current is shown in the top panel.  States 2a and 2b weaken over the measured density range while states 2c and 2d strengthen over the same range.  See Fig. \ref{C_transport} for labels of each RIQHE state.}
\label{RIQHE_ALL}
\end{figure}

\begin{figure}[t]
\includegraphics[width=3in]{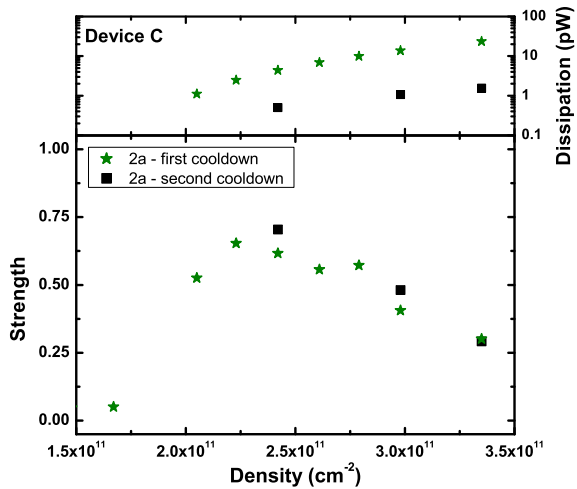}
\caption{(Color online). Comparison of the strength of the 2a RIQHE state in device C from two different cooldowns.  The strength of the state is comparable between the two cooldowns despite the large change in gate power dissipation.}
\label{RIQHE_2a}
\end{figure}

After optimizing our fabrication recipe, we cooled one exemplary device each from wafers B (device B) and C (device C) to low temperature ($< 25$ mK)  to examine the transport as a function of density at low temperature.  The RIQHE states in device C (from wafer C) showed an interesting evolution with density as shown in Fig.~\ref{RIQHE_ALL}.  In order to quantitatively compare the states, we defined the strength $S$ of the RIQHE states as 
\begin{equation}
S \equiv \frac{|R_{xy}^{c} - R_{xy}|}{|R_{xy}^{c}-R_{xy}^i|}
\end{equation}
where $R_{xy}^{c}$ is the the classical Hall resistance at the filling fraction of interest, $R_{xy}$ is the actual Hall resistance at the peak position, and $R_{xy}^i$ is the resistance of the nearest integer Hall plateau.  Using this definition, a fully quantized RIQHE state has a strength of 1 while a completely absent state has a strength of 0.  Figure \ref{RIQHE_ALL} shows the evolution of the RIQHE states in device C during its second cool-down.  The states on the high field side of $\nu = 5/2$, 2a and 2b ($\nu \sim$ 2.29 and 2.42, respectively), are seen to weaken over the measured density range while states 2c and 2d ($\nu \sim$ 2.56 and 2.70, respectively) continue to strengthen.  Figure \ref{RIQHE_2a} shows a comparison of the evolution of state 2a in device C as a function of density for two different cooldowns.  Even though the power dissipation from the gate leakage varied by $\sim$ 1 order of magnitude between the two cool-downs (possibly due to slightly different illumination conditions), the data show the same trend.  Comparisons between the other three states for the two cool-downs show similar agreement.  This appears to indicate that the observed evolution in strength is driven primarily by the 2DEG density and not by heating from the gate leakage.  At present, the origin of this behavior is not understood.  Regardless of the mechanism that causes states 2a and 2b to weaken with increasing density, this behavior is qualitatively different than that seen in states 2c and 2d and may point to a difference in the underlying localization mechanisms.  In contrast, the strength of all the RIQHE states in device B (from wafer B) with the larger superlattice setback (data not shown) were seen to increase with density up to a density (power dissipation) of $2.67 \times 10^{11}$ cm$^{-2}$ (6.4 pW) after which all the RIQHE states weakened.  While we cannot identify the mechanisms that alter localization, it appears that the different proximity of the superlattice to the 2DEG in wafers B and C has a significant impact.  

\begin{table}[t]
\caption{Summary of gate power dissipation in each device around the threshold of observable heating in the RIQHE states.  P$_1$ is defined as the power dissipation with the strongest measured RIQHE states and P$_2$ is defined as the power dissipation at which \emph{all} the RIQHE states are first observed to weaken with increasing density.}
\begin{tabular}{  l c c }
\hline
\hline
Device & P$_1$ (pW) & P$_2$ (pW) \\ \hline
A - Copper Header & 1.5  & 67 \\
B -  Copper Header &  6.4  & 114  \\ 
C -  Copper Header & 1.6  & 38  \\
\hline
\hline
\end{tabular}
\label{heating}
\end{table}

Finally, to estimate the maximum acceptable power dissipation, we summarize the measured data points on either side of the heating threshold at which \emph{all} the RIQHE states started to weaken with increasing density in Table \ref{heating}. Given that the maximum strength in device B was observed for a power dissipation of 6.4 pW and the RIQHE states in device C had begun to weaken by 38 pW, we set $\sim$ 10 pW as the upper bound on acceptable power dissipation from the gate in our devices given the setup of our cryostat.  This extremely low power level serves to highlight the necessity of minimizing the gate leakage in order to study the 2$^{\text{nd}}$ LL.

\section{Impact of Heterostructure Design on the FQHE states}
\begin{figure}[t]
\includegraphics[width=3in]{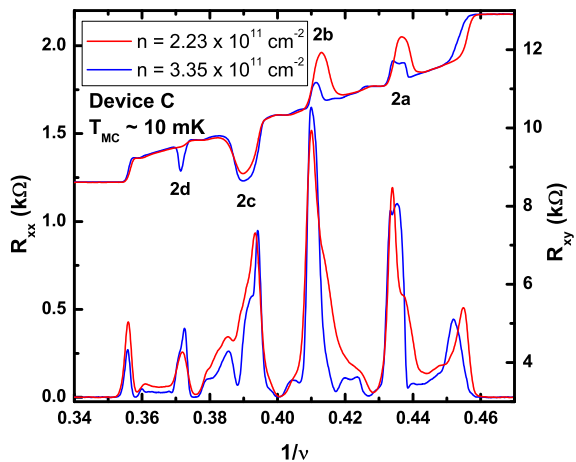}
\caption{(Color online). Magnetotransport in device C after illumination with a red LED.  The re-entrant states are labelled following the convention in reference \cite{deng2012collective}.  Red data show the transport for the maximum strength in RIQHE states 2a and 2b while the blue data show transport at the highest density before the second subband became occupied.}
\label{C_transport}
\end{figure}

We now turn to the central result of our work.  Figure \ref{C_transport} shows low temperature transport (T$_{\text{MC}}$ $\sim$ 10 mK) at two different densities for device C after illumination with a red LED. The device shows excellent transport with all four RIQHE states present and well developed FQHE states at $\nu$ = 14/5, 8/3, 5/2, and 7/3. In addition, nascent states at $\nu = 12/5$ and $\nu = 2 + 6/13$ begin to develop at high density.  This is, to our knowledge, the first time these states have been observed in a back-gated device, and their presence in spite of their extreme fragility \cite{xia2004electron, kumar2010nonconventional, zhang2012spin} further points to the high quality of the 2DEG.  The presence of the state at $\nu = 12/5$ is particularly interesting as this state has been proposed as a host of Fibonacci anyons which could be used for universal topological quantum computation \cite{nayak2008non-abelian}.

\begin{figure}[t]
\includegraphics[width=3in]{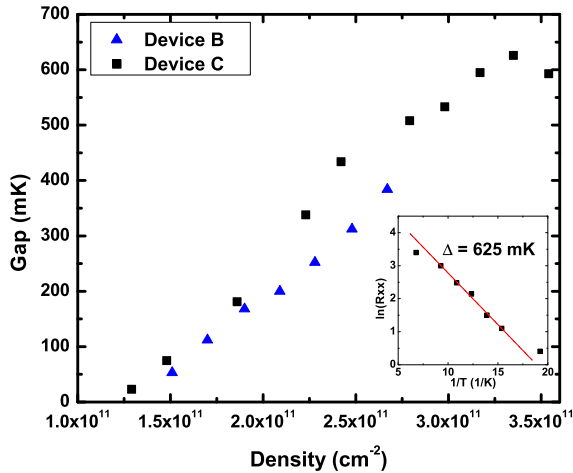}
\caption{(Color online). Gap at $\nu$ = 5/2 as a function of density for devices B and C.  Inset shows the Arrhenius plot for device C at a density of $3.35 \times 10^{11}$ cm$^{-2}$ where the gap was measured to be 625 mK.}
\label{Gaps}
\end{figure}

We examined the strength of the FQHE in each device quantitatively by measuring the gap at $\nu = 5/2$ ($\Delta_{5/2}$).  Figure \ref{Gaps} displays the gap at $\nu = 5/2$ as a function of density for devices B and C.  It is clear that, within the experimental resolution, the gaps are nearly identical for both devices at low density (n $< 2.5 \times 10^{11}$ cm$^{-2}$).  Evidently, neither the day-to-day variation in the MBE growth conditions nor the uncontrolled sample degradation from device fabrication nor the proximity of the superlattice to the 2DEG  significantly affect the gap at $\nu = 5/2$. Device C, however, allows investigation of much higher 2DEG densities.  Moreover, the magnitude of the gaps are very large with the gap in device C reaching a maximum value of 625 mK, the highest reported to date, at a density of $3.35 \times 10^{11}$ cm$^{-2}$.

\begin{figure}[t]
\includegraphics[width=3in]{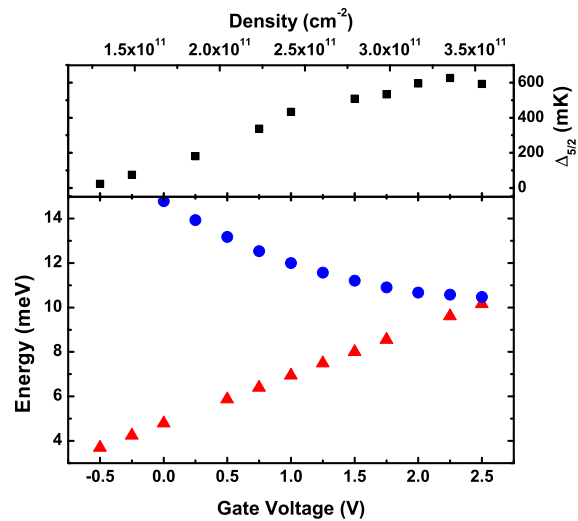}
\caption{(Color online). Cyclotron energy $\hbar\omega_c$ (red triangles) and spacing between E$_{\text{F}}$ and the second sub-band (blue circles) overlaid with $\Delta_{5/2}$ for Wafer C.  $\Delta_{5/2}$ drops suddenly at high density when the ground state is pushed into the lowest LL of the anti-symmetric sub-band.}
\label{Subband}
\end{figure}
One noticeable feature of the data from device C in Fig.~\ref{Gaps} is that at the highest density measured the gap shows a pronounced drop.  It has been previously reported \cite{liu2011anomalous} that the gap at $\nu = 5/2$ drops suddenly when the energy difference between the Fermi energy E$_{\text{F}}$ and the first excited electric sub-band in the quantum well equals the cyclotron energy. In this case, there is a level crossing and the ground state is pushed into the lowest LL of the anti-symmetric sub-band.  Figure \ref{Subband} shows the calculated \cite{nextnano} energy spacing along with the cyclotron energy as a function of density.  As expected, the experimentally measured gap at $\nu = 5/2$ is seen to drop suddenly when the cyclotron energy becomes approximately equal to the gap between E$_{\text{F}}$ and the second sub-band.  Taken together, our calculations and experimental data indicate that larger gaps at even higher densities could potentially be achieved if the quantum well were made more narrow to further separate the ground and excited sub-bands.

\section{Conclusion}

To summarize, we have examined the effect of heterostructure design and device processing on the performance of \emph{in-situ} back-gated 2DEGs in the 2$^{\text{nd}}$ LL.  We found that the position of the GaAs/AlAs superlattice barrier relative to quantum well has a large impact on the leakage characteristics of the device due to its effectiveness in blocking the diffusion of the Ohmic contacts towards the gate and minimizing Fowler-Nordheim tunneling. Moving the superlattice closer to the 2DEG greatly increases the range of low-leakage gating without significantly degrading the strength of the gap at $\nu = 5/2$ or other correlated states in the 2$^{\text{nd}}$ LL.  In addition, we found that gate leakage dissipation powers as small as a few pW are sufficient to cause electronic heating that impacts transport in the 2$^{\text{nd}}$ LL.  By improving the heat sinking of the lattice, the acceptable power dissipation is increased to $\sim$ 10 pW.  Moreover, it is likely that the FQHE gaps would continue to rise at higher density beyond what we report here if the electric sub-bands were spaced sufficiently far apart. Thus, examining gaps as a function of density in narrower quantum wells could potentially yield important results on the density dependence of the gap at other states potentially useful for topological quantum computation such as $\nu = 12/5$.  As we have demonstrated a robust recipe for these structures and as the FQHE states in the 2$^{\text{nd}}$ LL are very strong over a wide range of density, these devices should prove useful in experiments intended to test for the presence of non-Abelian statistics in quantum Hall systems.

\section*{ACKNOWLEDGEMENTS}
\vspace*{-10pt}
This work was supported by the US DOE Office of Basic Energy Sciences, Division of Materials Sciences and Engineering Award DE-SC0006671.  J.D.W.~thanks L.A. Tracy and R.L. Willett for helpful discussions regarding device processing.
%

%

\end{document}